\documentstyle[aps,prl,12pt,doublespace]{revtex}
\begin{document}

\draft


\title{Characterization of the Noise in Secondary Ion Mass Spectrometry
Depth Profiles}
\author{D.\ P.\ Chu, M.\ G.\ Dowsett, and G.\ A. Cooke}
\address{Department of Physics, University of Warwick, Coventry CV4 7AL, UK}

\maketitle

\begin{abstract}

The noise in the depth profiles of secondary ion mass spectrometry (SIMS)
is studied using different samples under various experimental conditions.
Despite the noise contributions from various parts of the dynamic SIMS
process, its overall character agrees very well with the Poissonian rather
than the Gaussian distribution in all circumstances.  The Poissonian
relation between the measured mean-square error (MSE) and mean can be used
to describe our data in the range of four orders.  The departure from this
relation at high counts is analyzed and found to be due to the saturation
of the channeltron used.  Once saturated, the detector was found to
exhibit hysteresis between rising and falling input flux and output
counts. 

\end{abstract}

\vskip 1cm


\newpage

\section{INTRODUCTION}

Secondary ion mass spectrometry (SIMS) is widely used for the measurement
of doping, impurity and matrix profiles in semiconductors. In this
application, the concentration data required may span $10$ orders of
magnitude overall, and $4$-$6$ orders for a particular species.  As the
SIMS data do not directly represent the true element profile, a data
quantification procedure is necessary.  Various methods from simple linear
mapping (ion dose to depth, signal to concentration) to deconvolution
using response functions of various types \cite{mgd941}.  Due to the
complexity of the dynamic SIMS profile process and the large record
dynamic range required, a very careful and unbiased treatment of the
measured SIMS data is vital to the success of a deconvolution method to be
used.  It is widely accepted that one of the best deconvolution methods 
in use is based on maximum entropy \cite{mgd942}.  We have developed and
applied a maximum entropy method to the deconvolution of SIMS profiles
\cite{mgd943,gac96}.  However, the use of the correct noise model is
essential in high dynamic range deconvolution, since the adoption of
arbitrary noise models can lead to quite different results -- both in the
depth resolution of the recoved profile, their dynamic range, and the
behaviour in the low counts regime.  Therefore, the accurate determination
of the noise characteristics is an indispensable part of the
self-consistent use of deconvolution methods. 

\section{NOISE IN SIMS DATA}

Noise in SIMS data can be attributed to four main sources: (1) the change
of the primary ion flux due to statistical fluctuations in the ion source;
(2) the fluctuation in the effective secondary ion yield due to the
statistical nature of sputtering; (3) the statistics of the secondary ion
detection process, probably dominated by ion to electron conversion
probabilities at the first collision with the detector (in the case of
this study, a Gallileo Corp. model 4816 channeltron); (4) there may also
be additional effects due to the locally random distribution of atoms of
the analytic species, significant when small volumes are analysed.  For
most depth profiles, integrating the signal over the volume of at least
$10^9\ {\rm nm}^3$ per data point makes this last contribution
insignificant.  In general, it is very difficult to distinguish the noise
generated by different sources within the same process.  The data we
normally collect and quantify in dynamic SIMS are the counts for a pixel
or frame, where the latter is effectively the sum over a set of the pixels
at a certain depth, in a period of time during which the operating
conditions of the instrument are assumed constant.  In other words, we can
only study the overall SIMS noise in our data. 

Separating noise from signal is not easy and convincing if the signal
itself is changing in time.  It is better to keep the signal as steady as
possible, {\it i.e.} to maintain the mean value of signal unchanged during
a period of time, and then deduce the noise from the collected data.  To
obtain a steady signal in dynamic SIMS profiling, we may use two different
approaches.  The first one is to study so-called staircase samples which
are made of different levels of constant doping of a specific element in
different ranges of depth.  Various levels of stable signals in the
corresponding mass channel will be obtained when we profile through the
samples.  The other approach is to collect data from a certain mass
channel which corresponds to a uniform distribution of an element while we
change the primary ion current from time to time.  Different currents will
generate different fluxes of secondary ions varying about a constant mean
flux. 

A thorough description of a noise requires its frequency distribution
\cite{rice54}.  If it differs from that of the noise model we wish to use,
the accuracy of the deconvolution process will be compromised. However,
there is no real problem in which we have sufficient information to fully
describe the frequency spectrum \cite{jaynes87}.  Fortunately, we do not
need so much information about the noise in our quantification procedure. 
Usually the relation between the magnitude of a noise, which is
characterized by its mean-square error (MSE), $\sigma^2 = \sum_i (s_i -
s_m)^2 / N$ where $s_i$ is the signal of the $i$-th frame out of a total
of $N$ frames, and the corresponding mean, $s_m = \sum_i s_i /N$, will be
sufficient in most cases. 

We would like to stress here that the relation between the MSEs and mean
values is not a full description of the noise, but rather a description of
our state of knowledge about the noise.  It is related to the fact that we
want to be fairly sure that we choose a set of possible solutions big
enough to include the true solution during our deconvolution process
\cite{jaynes87}.  Choosing too ``narrow'' a noise distribution will result
in data containing structure for which there is no evidence. 

\section{CHARACTERIZATION OF THE NOISE}

The samples used in this study were four boron staircase structures in
silicon grown by MBE at temperatures between $600$ and $760\ {\rm {}^\circ
C}$.  Doping levels varied from $3\times 10^{16}$ to $1\times 10^{19}\
{\rm atoms}/{\rm cm}^3$.  The samples are designated $14.01$, $17.03$,
$25.01$ and $39.02$ in what follows.  Depth profiles were obtained on our
EVA~3000 instrument using normally incident ${\rm O}^+_2$ ions at beam
energies of either $500\ {\rm eV}$ or $2\ {\rm keV}$ (typical of high
depth resolution conditions) using $600$ and $500\ {\rm {\mu}m}$ square
craters with $10\%$ areal gates (applied digitally, post acquisition).  A
typical boron profile from sample $39.02$ is shown in Fig.~\ref{fig1}. 
Two types of study were carried out: (i) profiles of the boron staircase
and silicon matrix intensity at fixed beam current, and (ii) profiles of
the silicon matrix channel at a variety of beam currents and sample bias,
$V_s$, to vary the input flux to the channeltron over a large dynamic
range.  In this way it was possible to establish a relationship between
$\sigma$ and $s_m$ over a wide range of samples and input fluxes.  The
experimental conditions are summarized in Tables~\ref{tab1} and
\ref{tab2} respectively. 

The $\sigma$ and $s_m$ of the noise calculated from the profiles of the
staircase samples study (i) are plotted in Fig.~\ref{fig2}, where the
$s_m$ were obtained from the segments of data in which the concentrations
are supposed to be constant.  In Fig.~\ref{fig2}, we can see that all the
$\sigma$ and $s_m$ agree well with the Poissonian relation $\sigma =
s_m^{1/2}$ rather than the Gaussian one $\sigma=(s_m/2)^{1/2}$ for $s_m
\le 10^4$. The data start to depart from this relation gradually when the
$s_m$ is larger.  Since the boron concentrations in our staircase samples
were low, we have the data points corresponding to the respective channels
in two spatially separated areas on the plot. 

Study (ii) determines whether or not the silicon has the same noise model
as the boron, despite the apparent departure from the Poissonian behaviour
shown in Fig.~\ref{fig2}. 

The corresponding $\sigma$ and $s_m$ are shown in Fig.~\ref{fig3}.  Again
we can see the Poissonian model suits our data well for $s_m \le 10^5$
with a departure for for larger input fluxes to the channeltron.  We
conclude therefore that under our experimental conditions the major noise
source or the combination of noise sources is effectively Poissonian. 

The MSEs and means were calculated from data of finite length.  We chose
long segments with large enough number of frames so that the statistical
errors involved are much smaller than the quantities in which we are
interested.  Some of the data exhibited a long term drift of a few \% in
intensity due to the normal behaviour of a plasma ion source.  To prevent
excessive MSE values from such data, MSE's were calculated from long
segments of data where each short segments had almost equal means with no
obvious systematic behaviour.  Some long segments consisted of a linear
drift on the mean, which was recognized by the linear relation of the
corresponding means for each short segment and the almost the same
magnitude of MSEs after subtracting a liner drift background.  In this
situation, the overall means were assigned to such MSEs.  This may lead to
a small fraction of error into the relation of the MSE and mean, but the
relative error involved is no more than the maximum relative difference of
the means of short segments within a long segment, which is negligible in
our calculations. 

To make sure that the data used in our calculations only involve random
noise, we used two methods to analyse them: time-frequency technique
\cite{lc95} and random walk method \cite{gmv96}. The first method uses
Wigner distribution $W(i,\omega)=(1 / {2\pi}) \int s^*(i-t/2) s(i+t/2)
\exp(-i\omega t) dt $ and is capable of revealing any ordering of
frequency components in time, and the second method calculates the root
mean square fluctuation of the displacement $F(j)=(<\Delta S(j)^2>-<\Delta
S(j)>^2)^{1/2}$ where $\Delta S(j)=S(j)-S(1)$ and the running sum
$S(j)=\sum_{i=1}^j s_i$ and can detect the presence of long-range
correlations existing in data.  No pattern with order was found in our
data and the $F(j) \sim j^\alpha$ gives $\alpha \rightarrow 1/2$ for large
$j$ as expected in uncorrelated time series. 

\section{HYSTERESIS EFFECT OF THE DETECTOR}

Note that the MSE of noise in Figs.~\ref{fig2} and \ref{fig3} becomes
graduately larger than what the Poissonian relation predicates as the
counts $>10^4$-$10^5$.  It appears only to be related to the input flux to
the detector.  For a better understanding of this phenomenon, we measured
the counts per frame in the silicon channel as a function of the primary
beam current, $I_p$.  See Figs.~\ref{fig4}(a) and (b).  We found that the
counts appear, as expected, to be proportional to the $I_p$ with
approximately the same slope when they are $< 10^4$ per second.  When the
output of the channeltron $> 10^4$ counts per second, the channeltron
starts to saturate.  When the $I_p$ is increased further, the output drops
as the pulse amplitude falls below the counting threshold.  We therefore
attribute the departures of the noise from the Poissonian relation in
Figs.~\ref{fig2} and \ref{fig3} to the channeltron saturation.  A simple
analysis shows that the MSE of counts saturates slower than the
corresponding mean as the channeltron approaches its saturation.  This
agrees qualitatively with the departure trends in both Figs.~\ref{fig2}
and \ref{fig3}. 

A hysteresis effect on the channeltron output was found as we raised the
$I_p$ to saturate the channeltron and then reduced it to the normal value. 
We see in Fig.~\ref{fig4}(b) that although the slopes for the linear
regimes before and after the channeltron is saturated are nearly the same,
the later output is about one order lower than the previous one for the
same input flux.  This may possibly be explained by degredation of the
surface of the channeltron capillary as a result of the large number of
secondary electrons flowing at the anode. This effect would make it very
difficult to quantify SIMS data if the channeltron is saturated at some
stages of profiling. 

\section{DISCUSSION AND CONCLUSION}

The Poissonian distribution is a result of ``law of small probability''
given by Poisson.  In SIMS, the ion count is much smaller than the number
of secondary ions ejected. This ratio, known as the useful yield, is $\sim
10^{-6}$-$10^{-2}$ for ${\rm B}$ and ${\rm Si}$ respectively on various
instruments \cite{ab87,jbc84}.  The ratio of counted ions to input primary
ions is likely to be orders of magnitude smaller still.  Such a small
probability will normally lead to a statistics of Poissonian type, and
several Poissonian stochastic processes combined together are {\it still}
Poissonian. 

We conclude that the noise in our dynamic SIMS depth profiles is
universally Poissonian despite of the change of samples and experimental
conditions.  Consequently, its MSE relates to the mean accurately as
described by the Poissonian distribution in all the linear dynamic range
of our experiments.  The departure from this relation is due to the
saturation effect of the channeltron used. 

It is worth mentioning that noise characterization is unconsciously or
deliberately ignored in many situations.  In fact, when one uses a certain
procedure of noise data processing, a particular distribution of noise is
usually assigned as asserting or hypothesizing a statement of fact.  The
one encountered the most is Gaussian noise, {\it e.g.} the Blackman-Tukey
window-smoothing procedure is associated with the hypothesis that the data
are a realization of a stationary Gaussian random process \cite{jaynes87}. 
Of course there is nothing to prevent one from applying a procedure to any
set of data no matter whether the hypothesis holds or not.  Moreover,
using a different type of noise may have no serious impact when one wants
merely to get a qualitative feature from the data studied.  However, use
of the wrong noise model is definitely not acceptable in forward methods
for deconvolution when the potential function contains noise-related
parameter(s).  Assigning a noise distribution different from real
situation can be very costly, as pointed out by Tukey {\it et al.}
\cite{tukey80}. 

There remain further investigations to be done for the noise in dynamic
SIMS profiles.  Changing the matrix composition may have effect on the MSE
and mean relation.  It will also be interesting to see whether such a
Poissonian relation still holds when the experimental condition is further
changed, such as using another element from a different type of source as
the primary ion. 

\section*{ACKNOWLEDGEMENT}

This work is supported by the EPSRC funding under the grant GR/K32715.

\newpage

\section*{Tables}

\begin{table}[hb]
\caption{Settings for the ${\rm B}$ staircase samples}
\begin{tabular}{ccc}
Sample/Run No. & Primary Beam & Primary Beam \\
{} & Energy $E_p$ (keV) & Current $I_p$ (nA) \\
\hline
14.01/1.1 & 0.5 & 243 \\
14.01/1.2 & 2.0 & 447 \\
17.03/1.2 & 0.5 & 243 \\
17.03/1.3 & 2.0 & 447 \\
17.03/1.4 & 2.0 & 390 \\
25.01/1.1 & 0.5 & 243 \\
25.01/1.2 & 2.0 & 447 \\
39.02/1.1 & 2.0 & 257 \\
39.02/1.2 & 2.0 & 469 \\
39.02/1.4 & 0.5 & 310 \\
39.02/1.4a& 0.5 & 310 \\
39.02/1.4b& 0.5 & 310 \\
\end{tabular}
\label{tab1}
\end{table}

\begin{table}[hb]
\caption{Settings for the ${\rm Si}$ only profiles}
\begin{tabular}{ccc}
Sample/Run No. & Primary Beam & Primary Beam  \\
{} & Energy $E_p$ (keV) & Current $I_p$ (nA) \\
\hline
14.01/2.1 & 0.5 & 0.014-133 \\
14.01/2.3 & 2.0 & 0.0095-120 \\
14.01/2.4 & 0.5 & 0.0064-260 \\
14.01/2.5 & 0.5 & 0.10-180 \\
14.01/2.6 & 0.5 & 0.01-160\\
\end{tabular}
\begin{tabular}{cccc}
Sample/Run No. & Frame Time & Areal & Sample Stage \\
{}        &  $t_f$ (sec)     & (\% crater) & Bias $V_{s}$ (V) \\
\hline
14.01/2.1 & 10 & 15 & 25.0 \\
14.01/2.3 & 1  & 15 & 20.0 \\
14.01/2.4 & 1  & 15 & 25.0 \\
14.01/2.5 & 1  & 100 & 25.0 \\
14.01/2.6 & 1  & 100 & 25.0\\
\end{tabular}
\label{tab2}
\end{table}

\newpage

\section*{Figure captions}

\begin{figure}[hbt]
\caption{
A typical SIMS profile for a boron staircases in silicon (Sample 
39.02, Run No: 1.1), obtained using $2.0\ {\rm keV}$ $^{16}$O$^+_2$
ion beam, current $260\ {\rm nA}$.  The four steps in the boron
channel between frames $200$ and $3500$ correspond to dopant levels 
of $3\times 10^{16}$, $1\times 10^{17}$, $1\times 10^{18}$ and
$1\times 10^{19}\ {\rm atoms}/{\rm cm}^3$ respectively.
}
\label{fig1}
\end{figure}

\begin{figure}[hbt]
\caption{
$\sigma$ vs $s_m$ for the noise in both boron and silicon mass channels 
of the staircase samples.  The dashed and dotted lines are for the 
Poissonian and Gaussian noise respectively. } 
\label{fig2}
\end{figure}

\begin{figure}[hbt]
\caption{
$\sigma$ vs $s_m$ for the noise in the silicon only profiles.  The 
dashed and dotted lines are corresponding to the relations of Poissonian 
and Gaussian noise respectively. }
\label{fig3}
\end{figure}

\begin{figure}[hbt]
\caption{
The counts of channeltron vs $I_p$.  The $I_p$ in (a) was increased
monotonically, while in (b) it was raised and reduced as shown by
the arrows.  The lines connecting the data points are drawn to guide 
eyes.}
\label{fig4}
\end{figure}

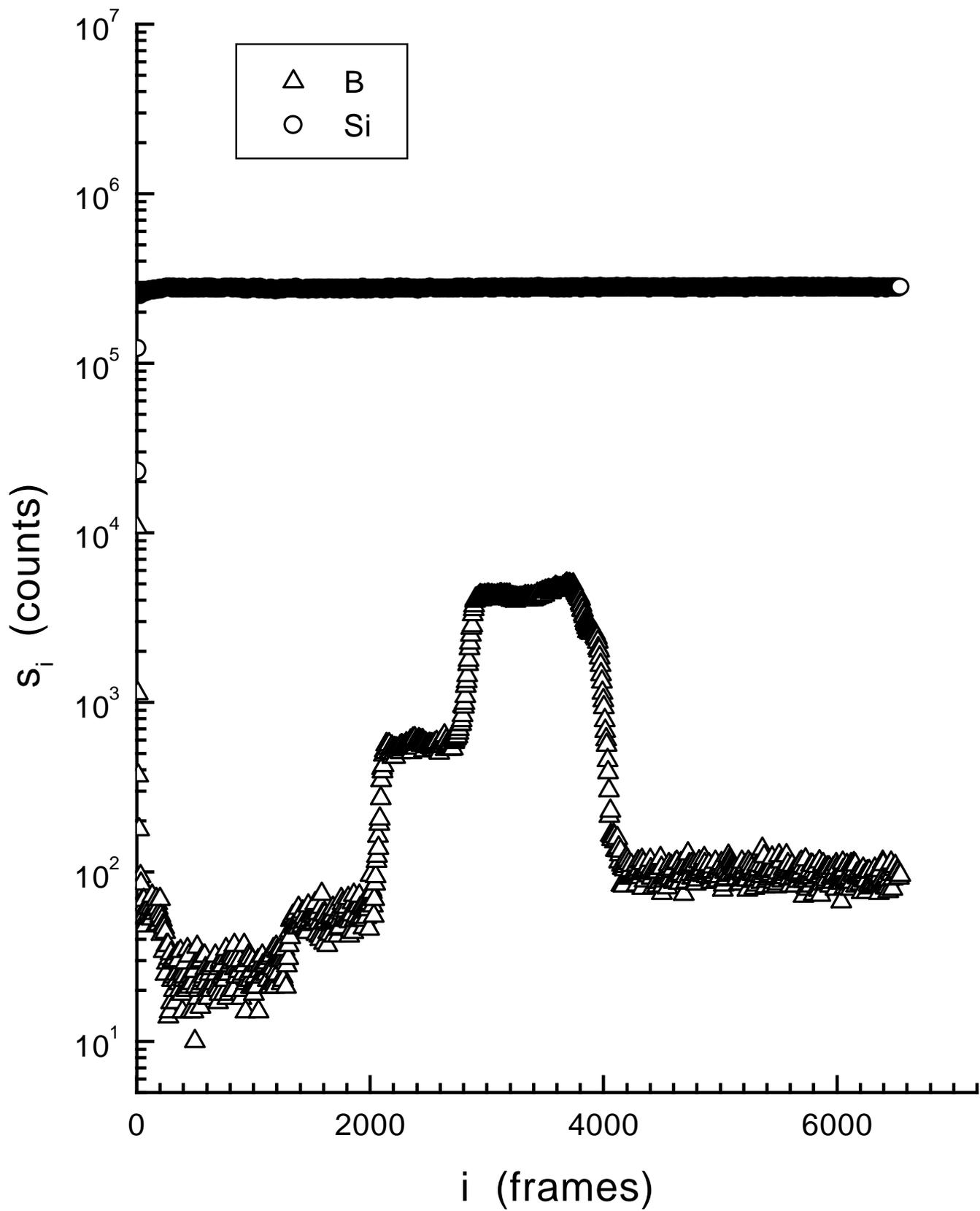
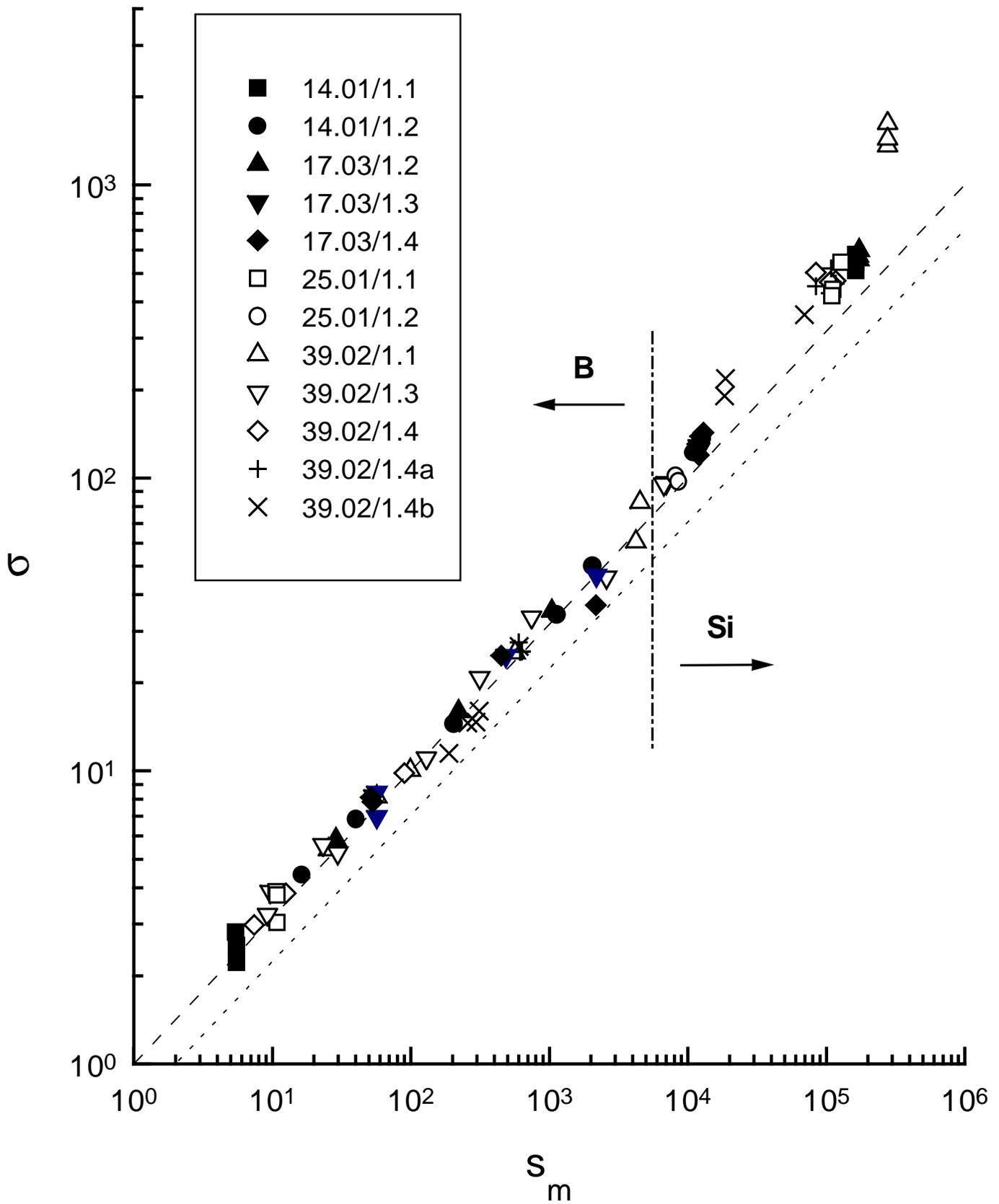
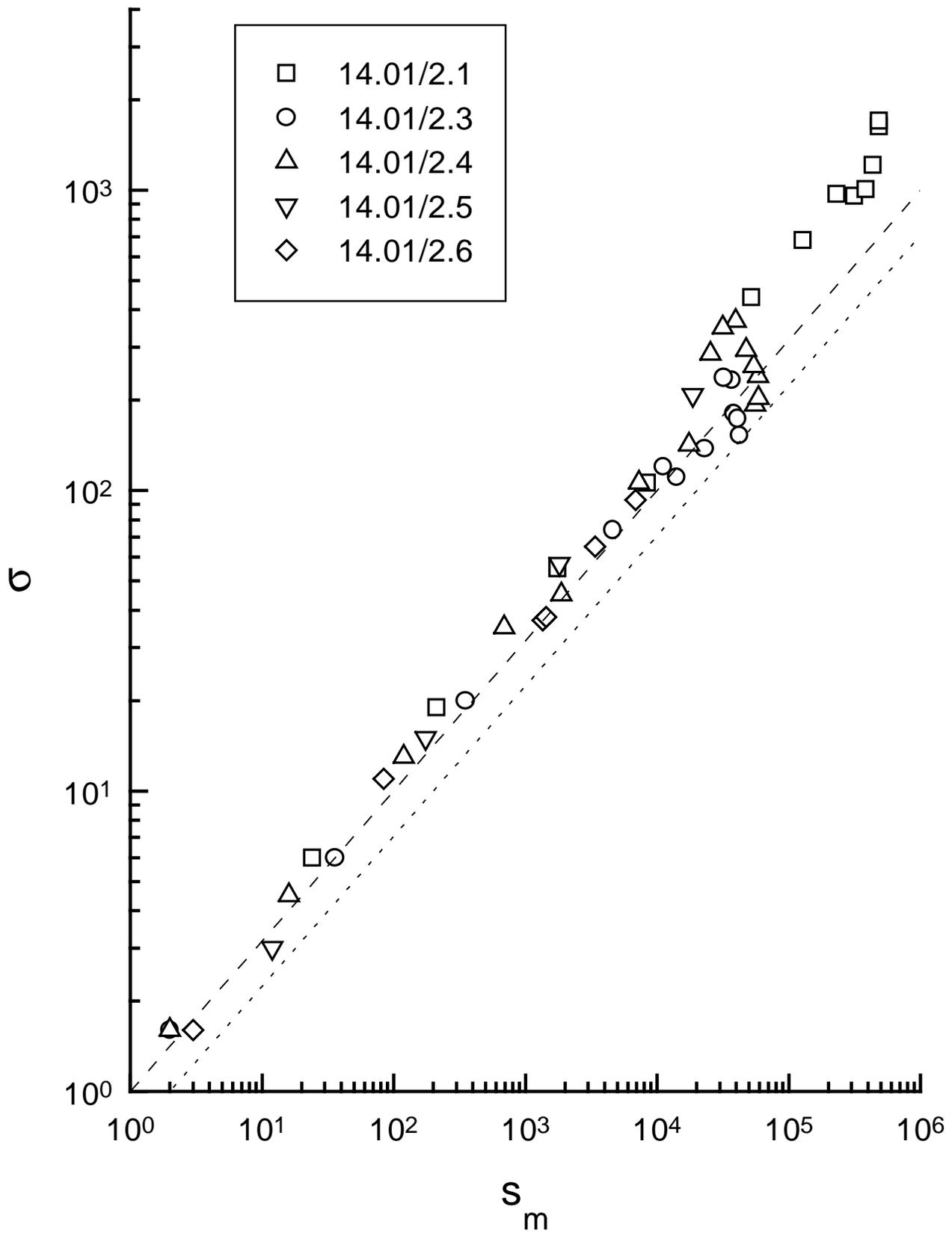
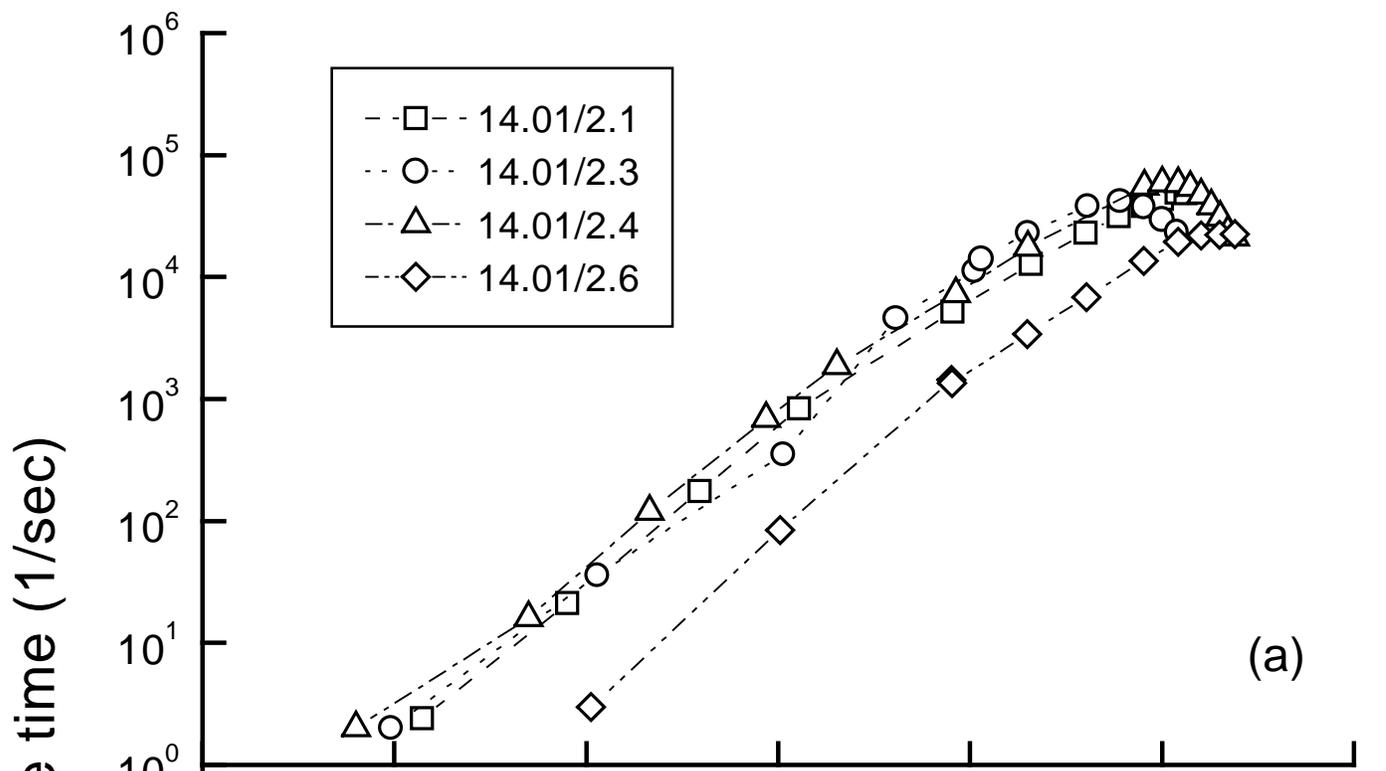
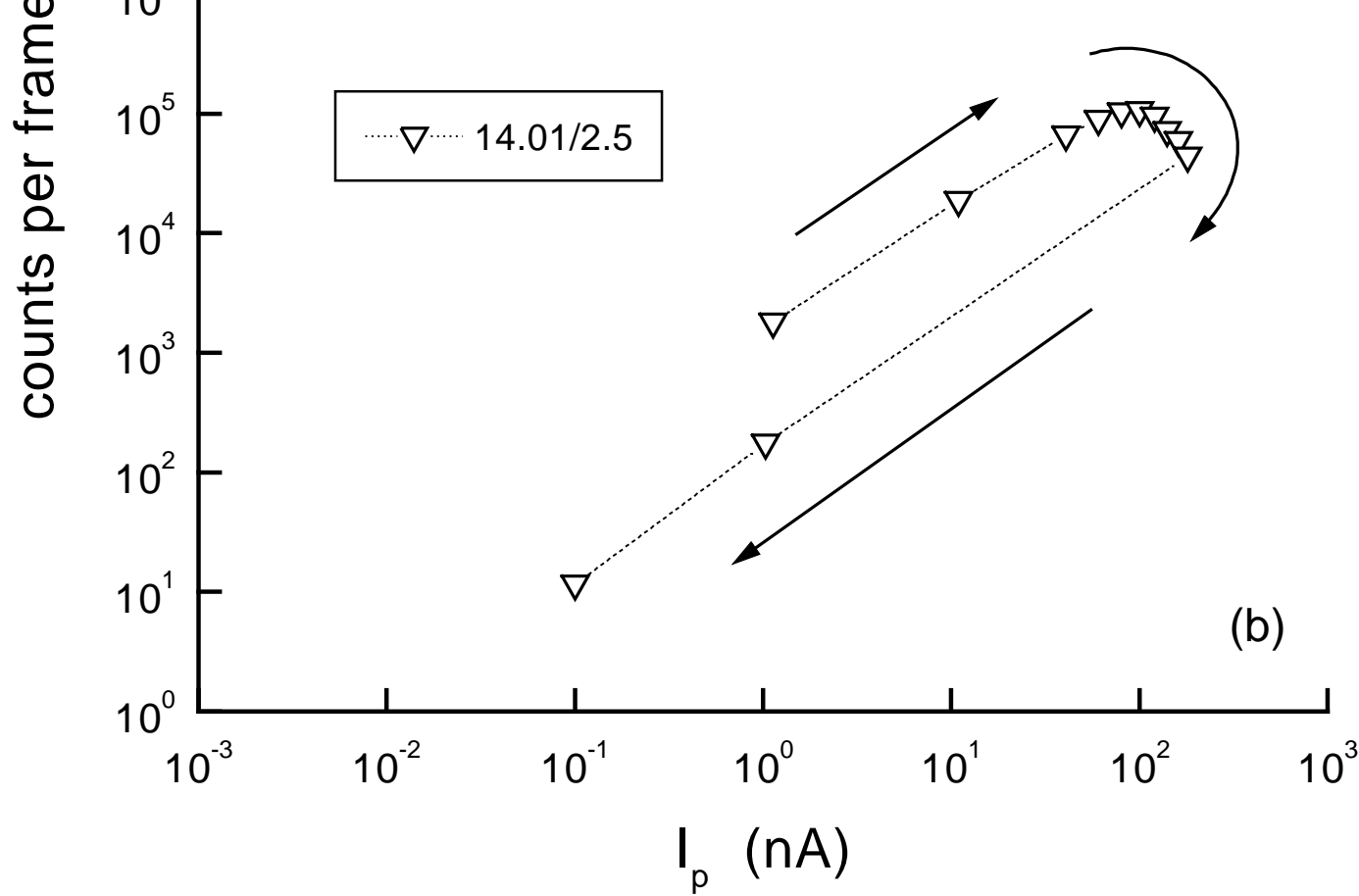
\end{document}